\documentclass[aps,prc,twocolumn,showpacs,superscriptaddress]{revtex4-1}

\usepackage{amsfonts,amsthm,latexsym}
\usepackage{graphicx}
\usepackage{bm}
\usepackage{xcolor}
\usepackage{soul} 
\usepackage{slashed}
\usepackage{mathrsfs}
\usepackage{latexsym}
\usepackage{longtable}
\usepackage{graphicx}
\usepackage{epsfig}
\usepackage{natbib}
\usepackage{amsmath}
\usepackage{bbm}
\usepackage{amssymb}
\usepackage{color}
\usepackage{hyperref}
\usepackage{url}
\usepackage{multirow}
\usepackage[utf8]{inputenc}

\usepackage{soul} \usepackage{subfigure} \usepackage{scalerel}
\usepackage{multirow} \AtBeginDocument{\usepackage{booktabs}}

\begin{document}


\title{Hybrid stars with color superconducting cores in an extended FCM model}

\author{Daniela Curin} \email{danielacurin@fcaglp.unlp.edu.ar}
\affiliation{Grupo de Gravitaci\'on, Astrof\'isica y Cosmolog\'ia,
  Facultad de Ciencias Astron\'omicas y Geof\'isicas, Universidad
  Nacional de La Plata, Paseo del Bosque S/N, La Plata (1900),
  Argentina.}

\author{Ignacio F. Ranea-Sandoval} \email{iranea@fcaglp.unlp.edu.ar}
\affiliation{Grupo de Gravitaci\'on, Astrof\'isica y Cosmolog\'ia,
  Facultad de Ciencias Astron\'omicas y Geof\'isicas, Universidad
  Nacional de La Plata, Paseo del Bosque S/N, La Plata (1900),
  Argentina.}  \affiliation{CONICET, Godoy Cruz 2290, Buenos Aires
  (1425), Argentina.}

\author{Mauro Mariani} \email{mmariani@fcaglp.unlp.edu.ar}
\affiliation{Grupo de Gravitaci\'on, Astrof\'isica y Cosmolog\'ia,
  Facultad de Ciencias Astron\'omicas y Geof\'isicas, Universidad
  Nacional de La Plata, Paseo del Bosque S/N, La Plata (1900),
  Argentina.}  \affiliation{CONICET, Godoy Cruz 2290, Buenos Aires
  (1425), Argentina.}

\author{Milva G. Orsaria} \email{morsaria@fcaglp.unlp.edu.ar}
\affiliation{Grupo de Gravitaci\'on, Astrof\'isica y Cosmolog\'ia,
  Facultad de Ciencias Astron\'omicas y Geof\'isicas, Universidad
  Nacional de La Plata, Paseo del Bosque S/N, La Plata (1900),
  Argentina.}  \affiliation{CONICET, Godoy Cruz 2290, Buenos Aires
  (1425), Argentina.}

\author{Fridolin Weber} \email{fweber@sdsu.edu, fweber@ucsd.edu}
\affiliation{Department of Physics, San Diego State University, 5500
  Campanile Drive, San Diego, CA 92182, USA.} \affiliation{Center for
    Astrophysics and Space Sciences, University of California at San
    Diego, La Jolla, CA 92093, USA.}

\date{1 October 2021} 

\begin{abstract}
We  investigate the  influence of repulsive vector interactions and
color superconductivity on the structure of neutron stars using an
extended version of the field correlator method (FCM) for the
description of quark matter. The hybrid equation of state is
constructed using the Maxwell description, which assumes a sharp
hadron-quark phase transition.  The equation of state of hadronic
matter is computed for a density-dependent relativistic lagrangian
treated in the mean-field approximation, with parameters given by the
SW4L nuclear model. This model described the interactions among
baryons in terms of $\sigma, \omega, \rho, \sigma^*$, and $\phi$
mesons.  Quark matter is assumed to be in either the CFL or the 2SC+s
color superconducting phase.  The possibility of sequential
(hadron-quark, quark-quark) transitions in ultra-dense matter is
investigated.  Observed data related to massive pulsars,
gravitational-wave events, and NICER are used to constrain the
parameters of the extended FCM model. The successful equations of
state are used to explore the mass-radius relationship, radii, and
tidal deformabilities of hybrid stars.  A special focus lies on
investigating consequences that slow or fast conversions of quark-hadron
matter have on the stability and the mass-radius relationship of
hybrid stars. We find that if slow conversion should occur, a new
branch of stable massive stars would exist whose members have radii
that are up to 1.5~km smaller than those of conventional neutron stars
of the same mass.  Such objects could be possible candidates for the
stellar high-mass object of the GW190425 binary system.
\end{abstract}


\maketitle{}

\section{Introduction} \label{intro}

Neutron stars (NSs) are compact stellar remnants which are born in
type-II supernova explosions~\cite{2004Sci...304..536L}. Within~just a
few minutes after birth, they turn into cold (on the nuclear scale)
stellar objects with temperature of just a few MeV~\cite{pons}. Their
masses can be as high as $\sim$ $2\, M_{\odot}$, and~their radii range
from $\sim$$10$ to $\sim$$13$~km, depending on mass.  The~mean density
of a NS with a canonical mass of $1.5 \, M_\odot$ is higher than the
nuclear saturation density of $n_0=2.5 \times 10^{14}$~g~cm$^{-3}$ and
the density reached in the central core is expected to be several
times higher than $n_0$ \cite{Potekhin_2010}.  Paired with the
unprecedented current progress in observational astronomy
~\cite{GW170817-new,Abbott:2018wiz,gw190425-detection,
  Riley:2021pdl,Miller:2021qha} these characteristic features make NSs
superb astrophysical laboratories for a wide range of physical
studies, which help us to understand the nature of matter subjected to
most extreme conditions of pressure and density
~\cite{Baym:2017whm,Orsaria:2019ftf}.

Traditionally, NSs are though as three-layer compact objects composed
by an inner core, an~outer core and a crust. Densities in the crust
are lower than $n_0$. Experimental nuclear physics data from
terrestrial laboratories have been extremely useful to reduce the
uncertainties in the low-density regime of such NS matter and its
associate equation of state (EoS).  The~situation is different for
matter with densities above $n_0$, for~which there is no general
agreement about the structure and composition.  This lack of knowledge
increases with increasing central density
~\cite{Weber:1999book,FW_qm}. Over~the years, several different
theoretical possibilities regarding the unknown nuclear composition
have been explored, including those that take into account a possible
hadron-quark phase transition (see, e.g.,
Refs.~\cite{FW_qm,Orsaria:2019ftf}, and~references therein). Neutron
stars containing hadrons and deconfined quarks in their center are
referred to as hybrid stars (HS).
The situation is different for matter at densities above $n_0$, for~which no general agreement on the structure and composition exists.
This lack of knowledge deepens with increasing density. Several
different theoretical possibilities of the central composition of NSs are
being explored, including some which account for a possible
hadron-quark phase transition (see, for~example,
Refs.~\cite{FW_qm,Orsaria:2019ftf}, and~references therein). NSs
containing hadrons and deconfined quarks in their centers, are known
as hybrid stars (HSs).

Observations of the $2\,M_\odot$ pulsars PSR J1614{-}2230~\cite{Demorest:2010bx,Arzoumanian_2018}, PSR J0348{+}0432~\cite{Antoniadis:2013pzd} and PSR J0740{+}6620~\cite{2020NatAs...4...72C} have imposed strong constraints to the EoS
of matter inside NSs. In~addition, the~merger of two binary NSs (BNSs)
known as GW170817, together with the detection of the electromagnetic
radiation associated with this event, has been used to put constraints
on the radius and dimensionless tidal deformability of the merging
compact objects and, therefore, indirectly on their nuclear EoSs
~\cite{GW170817-new}. The~analysis of the data from this event has been
used to set new limits on the radius of a $1.4\, M_\odot$ NS which is
estimated to be between $9.2$ and $13.76$~km
~\cite{Tews:2018ceo}. Moreover, the~upper bound to the maximum-mass of
cold and slowly rotating NS has been estimated to be $\sim$ $2.3\,
M_\odot$ \cite{NSMmax}. A~second BNS merger, named GW190425, has been
detected by the LIGO Livingston interferometer. In~this case, the~inferred total mass of the NSs that merged has been estimated to be
$M_{\rm tot} = 3.4^{+0.4}_{-0.1}~M_\odot$~\cite{gw190425-detection}. This is higher than the expected Galactic
mean mass for this kind of binary systems~\cite{BNS-mass}. To~date, no
electromagnetic counterpart associated with GW190425 has been detected
(see, for~example, Ref.~\cite{EM-counterpart2020MNRAS.497..726G}, and~references therein).

In 2019 the NICER collaborations have determined the mass and radius
of the isolated NS PSR J0030+0451 with values of
$1.34^{+0.15}_{-0.16}\, M_\odot$ and $12.71^{+1.14}_{-1.19}$~km~\cite{Riley2019} and $M=1.44^{+0.15}_{-0.14}\, M_\odot$ and
$13.02^{+1.24}_{-1.06}$~km~\cite{Miller2019}.  Very recently, data
from NICER and XMM-Newton were used to determine the radius of PSR
J0740+6620 with a value of $13.7^{+2.6}_{-1.5}$~km~\cite{Miller:2021qha} and $12.39^{+1.30}_{-0.98}$~km~\cite{Riley:2021pdl}. These values show that the radius of PSR
J0030+0451 is similar to the radius of the much more massive NS PSR
J0740+6620, whose mass is $2.072^{+0.067}_{-0.066}\, M_\odot$~\cite{Riley:2021pdl}. This constrains the nuclear EoS to
a greater degree than previously~possible.

For a comprehensive study of the properties of matter in the cores of
neutron stars and the EOS associated with such matter, it is necessary to resort to
Quantum Chromodynamics (QCD), the~theory of strong
interactions. Besides~ quark confinement,  asymptotic freedom is
one of the main features of QCD, which states that matter at high
density and/or temperatures exhibits a phase transition in which
hadrons merge leading  to the formation of a plasma of quarks
and gluons.  QCD has inherent computational problems that  make it
extremely difficult to perform analytic calculations at finite densities to be
performed. For~this reason, several phenomenological and/or effective
models have been proposed that reproduce (some of) the key features
and symmetries of the QCD Lagrangian density (see
Ref.~\cite{Orsaria:2019ftf} and references therein).

If the hadron-quark phase transition occurs in the cores of NSs, it
has been shown that the liberated quarks should form a color
superconductor (CSC)
\cite{Rajagopal:2001.WS,Alford:2001dt,Alford_2008}. This phase is
characterized by the formation of quark Cooper pairs, similarly to the
formation of electron Cooper pairs in ordinary condensed matter
superconductivity, which is energetically favored since it lowers the
energy of the Fermi sea of fermions~\cite{BCS_1957}.  A~Cooper pair of
quarks can not be a in a color singlet state as the corresponding
condensate breaks the QCD local color symmetry, SU(3)$_{\rm{color}}$. Hence
the notion color superconductivity.  Since the pairing among the
quarks is quite robust, quark matter, if~existing in the cores of NSs,
ought to be a CSC.  In~contrast to ordinary condensed matter
superconductivity, however, the~condensation patterns of CSC quark
matter are much more complex as up to three different quark flavors
and three different color states are involved in the diquark formation~\cite{Alford:2001dt,Alford_2008}.
  
Two of the most studied color superconducting phases are the two
flavor color superconducting (2SC) phase and the color-flavor-locked
(CFL) phase. In~the 2SC phase, only up, $u$, and~down, $d$, quarks
pair. The~strange quark, $s$, has a mass that is by two orders of
magnitude higher than the masses of $u$ and $d$ quarks. This favors
the formation of the 2SC phase at intermediate densities, while at
high densities, where the mass of the strange quark plays a less
dominant role, the~CFL phase may replace the 2SC phase. CFL is a more
symmetric phase of matter in which all three quark flavors are
involved in the pairing process. There is also the possibility that a
phase known as 2SC+s is formed at intermediate densities, where the
strange quarks are treated as a gas of free massive fermions~\cite{PhysRevC.96.065807}. The~formation of diquarks lower the energy
of the system by an amount related to the size of the CSC gap,
$\Delta$. This quantity is a function of the chemical potential, but~can be treated as a free parameter of the model~\cite{Lugones_2003}. This phenomenological approach is useful as it
gives theoretical insight into CSC.  The~occurrence of each of these
phases is directly related to the mass of the strange quark mass, the~energy gap, and~the electron chemical potential~\cite{Alford_2008}.

In addition to the possibility of diquarks formation in HSs, it is
known that the inclusion of the repulsive vector interaction in quark
models allows HSs to satisfy the $2\, M_{\odot}$ constraint~\cite{Orsaria:2014, Kl_hn_2015, ferreira2020quark}.

In this work, we study the influence of color superconductivity and of
vector interactions among quarks on the composition and structure of
HSs. Using an extended version of the Field Correlator Method (FCM)
for the description of quark matter~\cite{Simonov_2007b, Nefediev2009,
  Mariani:2016pcx}, the~effects of 2SC+s and CFL superconductivity is
included in the quark model in a phenomenological way. To~model the
hadronic phase of the hybrid EoS, we use the SW4L parametrization of
the density dependent relativistic mean-field theory which includes
all particles of the baryon octet as well as the $\Delta$ resonance~\cite{Malfatti2019hot}.

We assume that the surface tension at the hadron-quark interface is
high so that a sharp hadron-quark phase transition occurs, which is
modeled as a Maxwell transition (see Ref.~\cite{Orsaria:2019ftf}, and~references therein). In~this context we analyze the possibilities of
rapid versus slow conversions of matter at the hadronic and quark
matter interface~\cite{Pereira_2018}. This phenomenon requires a
modification of the traditional stability criteria of compact~objects.

The paper is organized as follows. In~Section~\ref{sec:2} we provide
some details of the treatment of phase transitions in HSs. Chemical
and mechanical equilibrium conditions for the construction of the
hybrid EoS are also given. Section~\ref{sec:hadronic} is devoted to
the description of the hadronic model used to describe the outer cores
of HSs. The~model used to describe quark matter in the inner core of
HSs is introduced in Section~\ref{sec:quark}. The~model accounts for
vector interactions among quarks and the effects of color
superconductivity. The~results of our comprehensive analysis of quark
matter parameters, phase transitions and hybrid configurations will be
discussed in Section~\ref{sec:res}. Finally, a~summary and discussion
of our key findings are presented in Section~\ref{sec:conclu}.

\section{Quark-Hadron Phase Transition in Neutron~Stars} \label{sec:2}

Properties such as the surface tension at the hadron-quark interface,
$\sigma _{\rm{HQ}}$, and~nucleation timescale are only poorly
known. These two quantities define the nature of the hadron-quark
phase transition. For~example, whether the hadron-quark phase
transition separating both types of matter is sharp or smooth is
determined by the value of the surface tension between the two
phases. If~the value of the surface tension is larger than a critical
value of $\sigma_{\rm HQ} \sim 70$~MeV~fm$^{-2}$, a~(sharp) Maxwell
phase transition is favored~\cite{Voskresensky:2002hu,Endo:2011em,Wu:2018zoe}. Otherwise, a~(smooth) Gibbs phase transition would be expected. It is important to
note that for the Gibbs formalism, the~global electric charge
neutrality condition leads to the appearance of geometrical structures
in the mixed hadron-quark phase. This so-called pasta phase is highly
dependent on the EoS used to construct the phase transition as well as
on the value of $\sigma _{\rm{HQ}}$ (see, for~example
Refs.~\cite{pasta:2019,pasta2019Univ....5..169W}, and~references
therein).

Although the analysis of data from GW170817 and its electromagnetic
counterpart led to the conclusion that high-mass NSs may be expected
to have quark matter in their inner cores~\cite{Annala:2019puf}, there
is no direct observational evidence of the occurrence of a
hadron-quark phase transition in the interior of such objects. In~this
work, we assume that the favored transition scenario is that of a sharp
hadron-quark phase~transition.

Within this theoretical framework, we study two different regimes
related to the nucleation timescales of the sharp phase transition:
the slow and the rapid conversion. The~importance of considering these
different theoretical scenarios has been introduced in
Ref.~\cite{Pereira_2018}. In~that work, the~authors showed the huge
impact these two types of conversions have on the structure and
stability of HSs against radial oscillations.  The~main result was
that when a slow conversion rate is considered to occur inside of a
HS, the~star continues to remain stable against radial oscillations
(i.e., the~fundamental radial mode remains real valued) even beyond
the gravitational mass peak, where the mass is decreasing with
increasing central energy density (for  details, see
Ref.~\cite{Mariani:2019}). This finding differs drastically from the
standard stability criterion established for compact stars whereupon
stability of stars against radial oscillations is only possible if the
mass is monotonically increasing with~density.

The concept of slow and rapid conversion is linked to the relationship
between two very different timescales. These are the nucleation
timescale, i.e.,~the characteristic time during which a hadron (quark)
fluid element is converted into quark (hadronic) matter, and~the
characteristic timescale of the oscillation of the fluid elements. As~to the latter, the~fluid elements located near the transition
interface oscillate to regions of larger (smaller) pressures as the
oscillation stretches and compresses the matter in the star. The~hadron-quark conversion is slow (rapid) if the nucleation timescale is
much larger (smaller) that the one associated with the oscillations at
the interface separating the two~phases.

The strong and weak interactions have times scales that differ from
each other by many orders of magnitude ($\tau _{\rm{strong}} \sim
10^{-23}$~s, $\tau_{\rm{weak}} \sim 10^{-8}$~s). For~this reason it
has been proposed that the hadron-quark deconfinement process ought to
consist of two separate steps: the formation of a virtual drop of
out-of-$\beta$-equilibrium quark matter that will subsequently reach
chemical equilibrium. The~characteristic time scale of this process is
related to the difference between the Gibbs free energies of
equilibrium and out-of-$\beta$-equilibrium quark matter (for a more
detailed discussion, see, for~example,
Ref.~\cite{Bombaci:2007eoc}). Present results for this energy
difference are strongly model dependent and inconclusive (for details,
see, for~example,
Refs.~\cite{Haenseletal1989,Bombaci-Parenti-Vidana2004,
  Bombacietal2009, Lugones-Grunfeld2011,Bombacietal2016}). Therefore,
in this work we shall account for both theoretical possibilities and
analyze the astrophysical consequences and observational differences
that might help understand in detail the microphysics of the
hadron-quark phase~transition.

The composition of the matter in the interior of a HS is determined by
the condition of $\beta$-equilibrium and electric charge neutrality~\cite{1985ApJGlen,Alford_2002}.  These condition imposes a
relationship between the chemical potentials of the different particle
species in the hadronic phase,
\begin{equation}
\mu_{B} = \mu_n + q_{B}\,\mu_e \, ,
\label{chempot}
\end{equation}
and in the quark phase with flavors $f$ = $u$, $d$, $s$,
\begin{equation}
\mu_{f} = \mu_n/3 + q_{f}\,\mu_e \, ,
\label{chempot2}
\end{equation}
where $q_{B}$ and $q_f$ are the baryon and quark electric charges,
$\mu_n$ is the neutron chemical potential, and~$\mu_e$ the electron
chemical~potential.

To calculate the hybrid EoS within the Maxwell construction at zero
temperature, $T=0$, we impose the mechanical equilibrium condition that
reads
\begin{equation}
    P_h(\mu_n,\mu_e) = P_q(\mu_n,\mu_e) \, .
\end{equation}

Charge neutrality is imposed locally in the Maxwell formalism, i.e.,
each phase has to be independently electrically neutral. This
condition is satisfied if \mbox{$\partial P_{h(q)} / \partial \mu_e =
  0$}, where $P_{h(q)}$ is the pressure of the hadronic (quark) phase,
which will be defined~later.


\section{The Hadronic~Phase} \label{sec:hadronic}

To describe hadronic matter in the outer core of HSs we use the SW4L
parametrization of the density dependent non-linear relativistic
mean-field model~\cite{Typel:1999yq, Spinella2017:thesis,
  Malfatti2020PRD}.  This family of models have gained popularity
since the density-dependent couplings allows one to account for the
latest slope values of the symmetry energy consistent with
experimental data~\cite{Lattimer_2013, Lattimer2019}.  This quantity
plays a significant role for the determination of the radii of NSs~\cite{Horowitz_2014}. 

For the SW4L parametrization, the~interactions between baryons are
described by the exchange of scalar ($\sigma$, $\sigma^*$), vector
($\omega$, $\phi$) and isovector ($\rho$) mesons. The~pressure and the
energy density of the model are given by
\begin{eqnarray}\label{HM:pressure}
P_h &=& \frac{1}{\pi^2}\sum_B \int^{p_{F_B}}_0 \! dp \, 
\frac{p^4}{\sqrt{p^2+m_B^{*2}}}-\frac{1}{2} m_{\sigma}^2
\bar{\sigma}^2\nonumber\\ &-& \frac{1}{2} m_{\sigma^*}^2
\bar{\sigma^*}^2 + \frac{1}{2} m_{\omega}^2 \bar{\omega}^2 +
\frac{1}{2} m_{\rho}^2 \bar{\rho}^2+ \frac{1}{2} m_{\phi}^2
\bar{\phi}^2\\ &-& \frac{1}{3} \tilde{b}_{\sigma} m_N (g_{\sigma N}
\bar{\sigma})^3 - \frac{1}{4} \tilde{c}_{\sigma} (g_{\sigma N}
\bar{\sigma})^4 + n \widetilde{R} \,,  \nonumber
\end{eqnarray}
\begin{eqnarray}\label{HM:energy}
\varepsilon_h &=& \frac{1}{\pi^2}\sum_B \int^{p_{F_B}}_0 \! dp \, 
\sqrt{p^2+m_B^{*2}}+\frac{1}{2} m_{\sigma}^2
\bar{\sigma}^2\nonumber\\ &+& \frac{1}{2} m_{\sigma^*}^2
\bar{\sigma^*}^2 + \frac{1}{2} m_{\omega}^2 \bar{\omega}^2 +
\frac{1}{2} m_{\rho}^2 \bar{\rho}^2+ \frac{1}{2} m_{\phi}^2
\bar{\phi}^2\\ &+& \frac{1}{3} \tilde{b}_{\sigma} m_N (g_{\sigma N}
\bar{\sigma})^3 + \frac{1}{4} \tilde{c}_{\sigma} (g_{\sigma N}
\bar{\sigma})^4 \, ,  \nonumber
\end{eqnarray}
where the sum over $B$ sums all members of the baryon octet, $p, n,
\Lambda, \Sigma, \Xi$, as~well as the $\Delta$ resonance. The~quantities $g_{\rho B}(n)$ denote density dependent meson--baryon
coupling constants that have a functional form given by
\begin{equation}
g_{\rho B}(n) = g_{\rho B}(n_0)\, \mathrm{exp}\left[\,-a_{\rho}
  \left(\frac{n}{n_0} - 1\right)\,\right] \, ,
\end{equation}
where $n$ is the total baryon number density. The~last term in
Equation~(\ref{HM:pressure}) is the rearrangement term which guarantees the
thermodynamic consistency of the model~\cite{Hofmann:2001},
\begin{equation}
\widetilde{R} = [\partial g_{\rho B}(n)/\partial n]
    I_{3B} n_B \bar{\rho} \, .
\end{equation}    

The quantity $I_{3B}$ is the 3-component of isospin, and~\mbox{$n_{B}=p_{F_B}^3/3\pi^2$} are the particle number densities of
each baryon $B$ with Fermi momentum $p_{F_B}$. The~effective baryon
mass in Equations~(\ref{HM:pressure}) and (\ref{HM:energy}) is given by
\begin{equation}
m_B^*= m_B - g_{\sigma B}\bar{\sigma}-g_{\sigma^* B}\bar{\sigma^*} \,
.
\end{equation}

The parameters of SW4L are presented in
Table~\ref{table:parametrizations}. These values are adjusted to the
properties of nuclear matter at saturation density shown in
Table~\ref{table:properties} (for details, see
Ref.~\cite{Malfatti2020PRD}, and~references therein).

\begin{table}[htb]
\begin{center}
\begin{tabular}{cc}
\toprule 
Quantity ~~~~~~~~~~ &Numerical Value\\ 
\midrule
$m_{\sigma}$~(GeV)    & 0.5500          \\
$m_{\omega}$~(GeV)          &0.7826    \\
$m_{\rho}$~(GeV)          & 0.7753         \\
$m_{\sigma^*}$~(GeV)          & 0.9900         \\
$m_{\phi}$~(GeV)          & 1.0195         \\
$g_{\sigma N}$             & 9.8100         \\
$g_{\omega N}$            & 10.3906           \\
$g_{\rho N}$            & 7.8184           \\
$g_{\sigma^* N}$            & 1.0000           \\
$g_{\phi N}$            & 1.0000           \\
$\tilde{b}_{\sigma}$         & 0.0041                     \\
$\tilde{c}_{\sigma}$         & -0.0038                \\
$a_{\rho}$         & 0.4703            \\
\bottomrule
\end{tabular}
  \caption{Parameters of the SW4L parametrization that lead to the
    properties of symmetric nuclear matter at saturation density shown
    in Table~\ref{table:properties}.}
\label{table:parametrizations}
\end{center}
\end{table}

\begin{table}[htb]
\begin{center}
\begin{tabular}{cc}
\toprule 
Saturation Properties ~~~~~& Numerical Values\\
\midrule
$n_0$~(fm$^{-3}$)     & 0.15          \\
$E_0$~(MeV)          & -16.0     \\
$K_0$~(MeV)          & 250.0          \\
$ m^*_N/m_N$        & 0.7        \\
$J_0$~(MeV)          & 30.3           \\
$L_0$~(MeV)          & 46.5         \\
\bottomrule
\end{tabular}
  \caption{Energy per
    nucleon $E_0$, nuclear compressibility $K_0$, effective nucleon
    mass $m^*$, symmetry energy $J_0$, and slope of the symmetry energy
    $L_0$ of nuclear matter at saturation density, $n_0$,
    obtained for the SW4L parametrization.}
\label{table:properties}
\end{center}
\end{table}


\section{The Quark~Phase} \label{sec:quark}

To describe quark matter in the inner core of cold HSs we use an
extended version of the FCM model, including the effects of repulsive
vector interactions among quarks and of color~superconductivity.

The FCM model is based on the calculation of the amplitudes of the
color electric $D^E(x)$, $D_1^E(x)$ and color magnetic $D^H(x)$,
$D_1^H(x)$ Gaussian correlators. $D^E(x)$ and $D^H(x)$ are directly
related with the confinement of quarks, and~$D_1^E(x)$, $D_1^H(x)$
contain perturbative terms related to the perturbation expansion
over the strong coupling constant at a given order
~\cite{Simonov2007a,Simonov_2007b}. The~method has been generalized to
finite temperature and baryonic density using the single line
approximation (SLA) which neglects, to~first order, all perturbative
and non-perturbative interactions of the system. In~this way, it is
possible to factorize the partition function into the products of one
gluon and one quark (anti-quark) contributions and thus calculate the
corresponding thermodynamic potential~\cite{Nefediev2009}.

For zero-temperature HS matter, $D^E(x)=D^H(x)$ and
$D_1^E(x)=D_1^H(x)$, leaving two field correlators which can be
parametrized through the large distance $q\bar{q}$ potential, $V_1$,
and the gluon condensate, $G_{2}$. In~addition, the~main consequence
of repulsive vector interactions for HSs is to stiffen the EoS of
quark matter to obtain $2\, M_{\odot}$ stellar configurations, in~agreement with recent observations of massive pulsars. We also mention
that the onset of quark matter in the interior of HSs is affected by
this~interaction.

Both vector interactions among quarks and color superconductivity are
taken into account by FCM~model.
  
\subsection{Inclusion of Vector Interactions in the FCM~Model}  \label{sub:iv-FCM}

The inclusion of vector interactions among quarks modifies the SLA of
the quark pressure in the following way

\vspace{-24pt}

\begin{eqnarray}
P_f & =& \frac{T^4}{\pi^2}\,\left[\phi_\nu^+ (\frac{\mu_f^* -
    V_1/2}{T})+ \phi_\nu^- (\frac{\mu_f^* + V_1/2}{T}) \right]
\nonumber\\ &&+~P_{VI}(T, \mu_f^*) \, ,
\label{quark_press}
\end{eqnarray}
where
\begin{equation}
\phi_\nu^{\pm} (a) = \int_0^\infty dz \,
\frac{z^4}{\sqrt{z^2+\nu^2}}\,\frac{1}{e^{\sqrt{z^2 + \nu^2} \pm a} +
  1} \, ,
\end{equation}
and $\nu=m_f/T$, $m_q$ is the bare quark mass of a quark flavor $f$
and $T$ is the temperature. The~effective chemical potential is given
by
\begin{equation}
\mu^*_f=\mu_f-K_{\rm v}\,w(T,\mu_f^*) \, ,
\label{chem_eff}
\end{equation}
where $\mu_f$ is the chemical potential of a quark of flavor $f$,
$K_{\rm v}$ is the coupling constant of the vector interactions, and~$w(T,\mu_f^*)$ is the associated~condensate.

An expression similar to the first term in Equation~(\ref{quark_press}) can
be deduced for the pressure of the gluons, which vanished at zero
temperature. The~second term is the pressure due the vector
condensates given by
\begin{equation}
P_{VI}(T, \mu^*_f) = \frac{K_{\rm v}}{2}\,w^2(T,\mu_f^*) \, .
\end{equation}

Taking the limit $T \rightarrow 0$ in Equation~(\ref{quark_press}), we
obtain a simplified expression for the total pressure of the system that reads
\begin{eqnarray}
\nonumber P_q&=&\sum_{f=u,d,s}P_f=\sum_{f=u,d,s}\left[
  \frac{3}{\pi^{2}}\,\int^{p^*_F}_0\,z^{2} (\tilde{\mu}^*_f-z)
  \,dz\right. + \\ &&\left.+\frac{K_{\rm v}}{2}\,w_f^2\right] + \Delta
\epsilon_{vac} \, ,
 \label{zero_press}
\end{eqnarray}
where $\tilde{\mu}_f^*=\mu_f^*-V_1/2$,
$p^*_F=\sqrt{\tilde{\mu}^*_f-m_f^2}$, $w_f=w(\tilde{\mu}_f^*)$, and~\begin{equation}
\Delta \epsilon_{vac}=-\frac{11-\frac{2}{3}\,N_f}{32}\,\frac{G_2}{2} 
\end{equation}
is the vacuum energy density for $N_f$ flavors~\cite{Simonov_2007b}.

The EoS of the system can be computed using the Euler thermodynamic relation
given~by
\begin{eqnarray}
  \varepsilon= -P_{q} + \sum_{f=u,d,s} \mu_f
  \frac{\partial{P_f}}{\partial \mu_f}\, .
  \label{euler}
\end{eqnarray}

The effective chemical potential of Equation~(\ref{chem_eff}) is determined
in a self-consistent way by minimizing Equation~(\ref{zero_press}) with respect to the vector condensate, from~which it follows that
$\mbox{$w_f=n(\mu_f^*)$}$, where $n(\mu_f^*)$ is the number density
quark flavor $f$.

\section{Results}\label{sec:res}

The hybrid configurations studied in this work consist of an inner
core, an~outer core, and~a crust. The~latter has been modeled in out
study by the Baym-Pethick-Sutherland (BPS) and Baym-Bethe-Pethick
(BBP) EoSs~\cite{Baym1971tgs, Baym1971nsm}.

The FCM model has already been used in several works to model the
inner cores of HSs~\cite{plumari2013quark, logoteta2013quark,
  burgio2016hybrid, Mariani:2016pcx, Mariani:2019,
  khanmohamadi2020structure}. In~these studies, the~parameter space
($V_1$, $G_2$) of the model has been analyzed by accounting for
constraints from Lattice QCD simulations, the~existence~of $2~M_{\odot}$ pulsars, as~well as the limits set by the gravitational-wave
event GW170817. In~our work, we expand the ($V_1$, $G_2$) space by
accounting for vector interactions and color superconductivity, which
introduces the additional parameters $K_{\rm v}$ and $\Delta$,
respectively. To~investigate this new parameter space spanned by
$V_1$, $G_2$, $K_{\rm v}$, $\Delta$, we have chosen $V_1=20$~MeV and
$G_2=0.009$~GeV$^4$, following the results presented in
  Ref.~\cite{Mariani:2016pcx}; these values for $V_1$ and $G_2$
  are qualitatively representative of the parameters space. In~this
way, we focus our attention on the values of $K_{\rm v}$ and $\Delta$.

{In this context,} it should be mentioned that one of the
  methods used to combine and analyze different sets of data is
  Bayesian analysis, which analyzes the ranges of parameters using
  probability techniques.  {The application of Bayesian methods is
    frequently used in astrophysics (e.g., neutron star physics~\cite{char2020bayesian,xie2020bayesian}), particularly when
    dealing with large data sets.  A~Bayesian analysis of the
    parameters of our model, however, is out of the scope of this paper.} 

To calculate the properties of HSs, such as gravitational mass,
radius, tidal deformability and study their stability under slow and
rapid conversion of hadronic matter to quark matter, we solve the
relativistic hydrostatic equilibrium equation of Tolman, Oppenheimer,
and Volkoff (TOV) \cite{Tolman:1939jz, Oppenheimer:1039omn}.

\subsection{Analysis of the FCM Parameter Space Spanned by $V_1$, $G_2$, $K_{\rm~v}$, $\Delta$}

\begin{figure}
 \includegraphics[width=0.9\columnwidth]{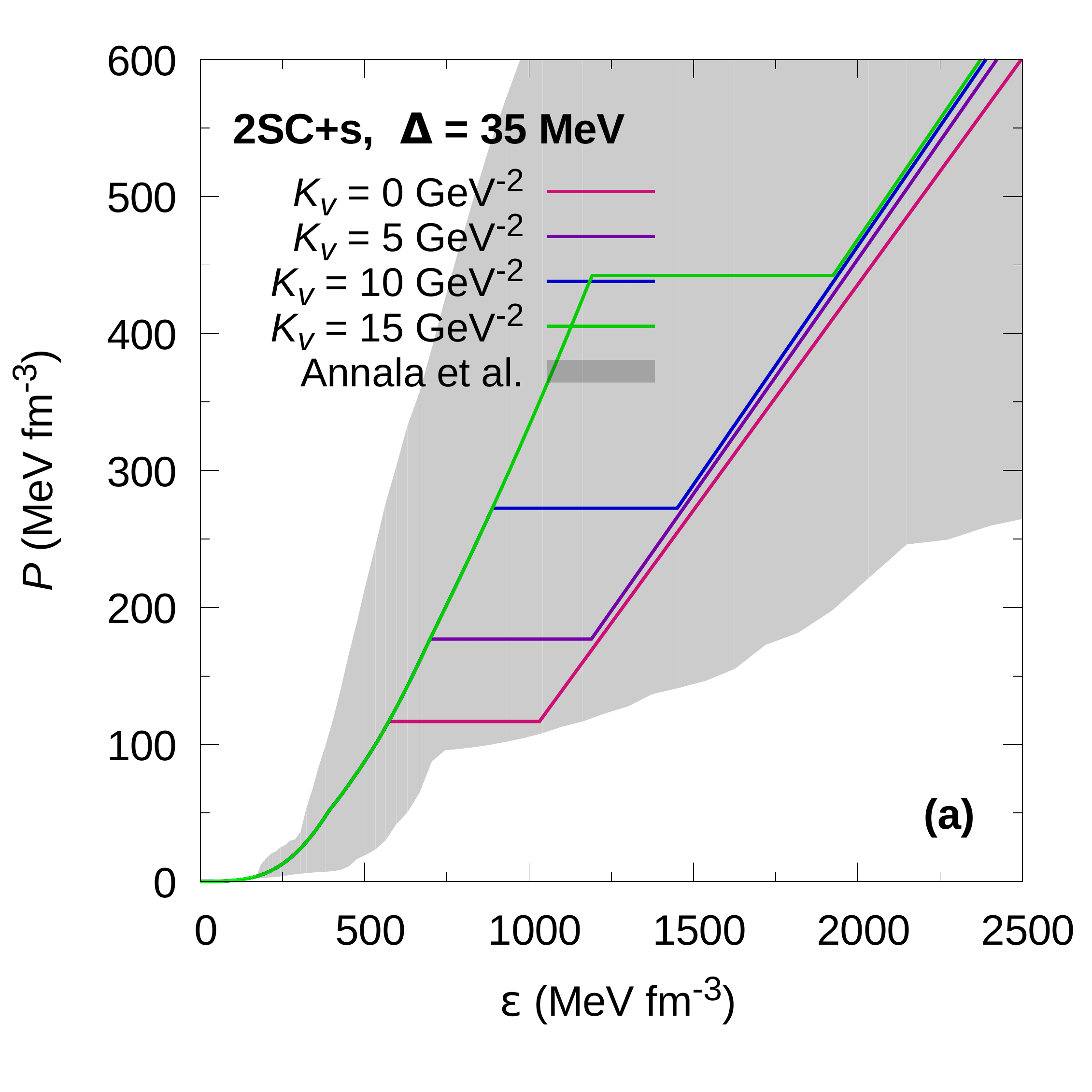}
 \includegraphics[width=0.9\columnwidth]{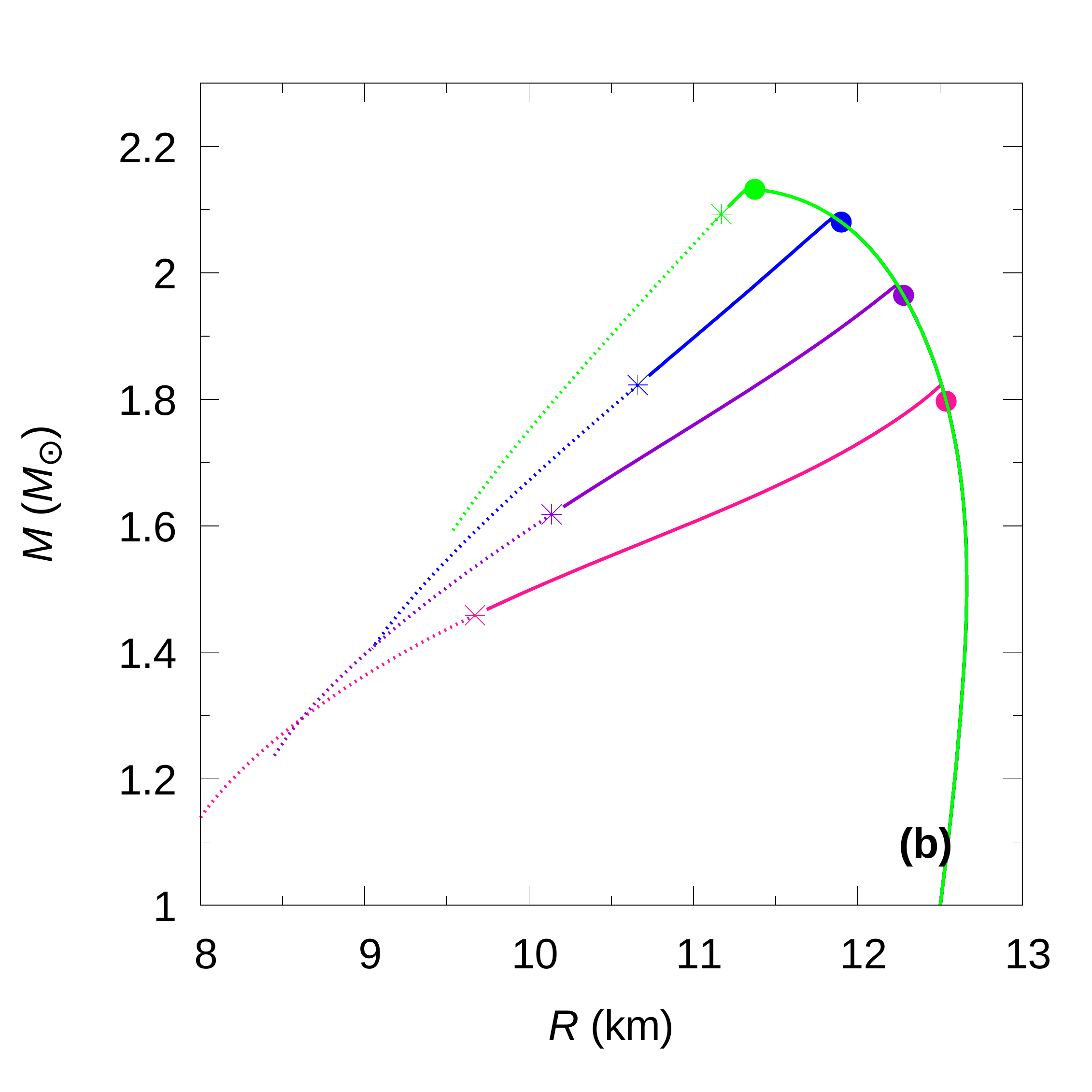}
 \caption{{(Color online) Hybrid EoS} (panel (\textbf{a})) and mass-radius
   relationship (panel (\textbf{b})) for the 2SC+s phase at fixed gap value of
   $\Delta=35$~MeV, for~different values of the $K_{\rm v}$ parameter.
   In panel (\textbf{a}), the~grey region shows the constraints presented
     in~\cite{Annala:2019puf}.  The~solid dots in panel (\textbf{b}) indicate
   the appearance of the color superconducting phase, just before the
   maximum mass peak. For~rapid conversions, the~stellar
   configurations to the left of each maximum mass are unstable and
   the existence of HSs is only marginal. For~slow conversions, an~   extended stability branch exists. The~stable configurations are
   shown by continuous lines. The~terminal configurations are marked
   with asterisks.} 
 \label{fig:2sc_delta35}
\end{figure}

\begin{figure}
  \includegraphics[width=0.9\columnwidth]{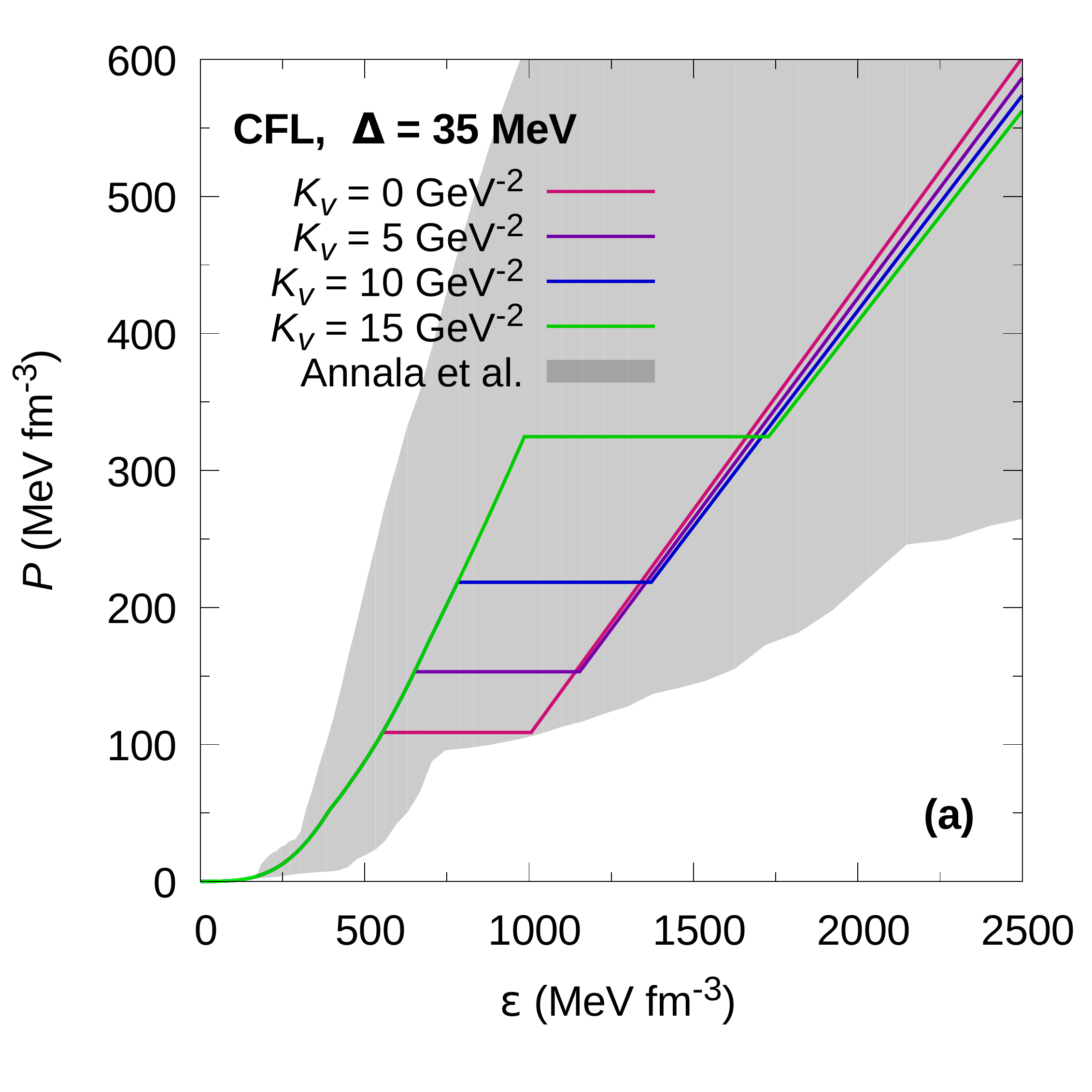}
  \includegraphics[width=0.9\columnwidth]{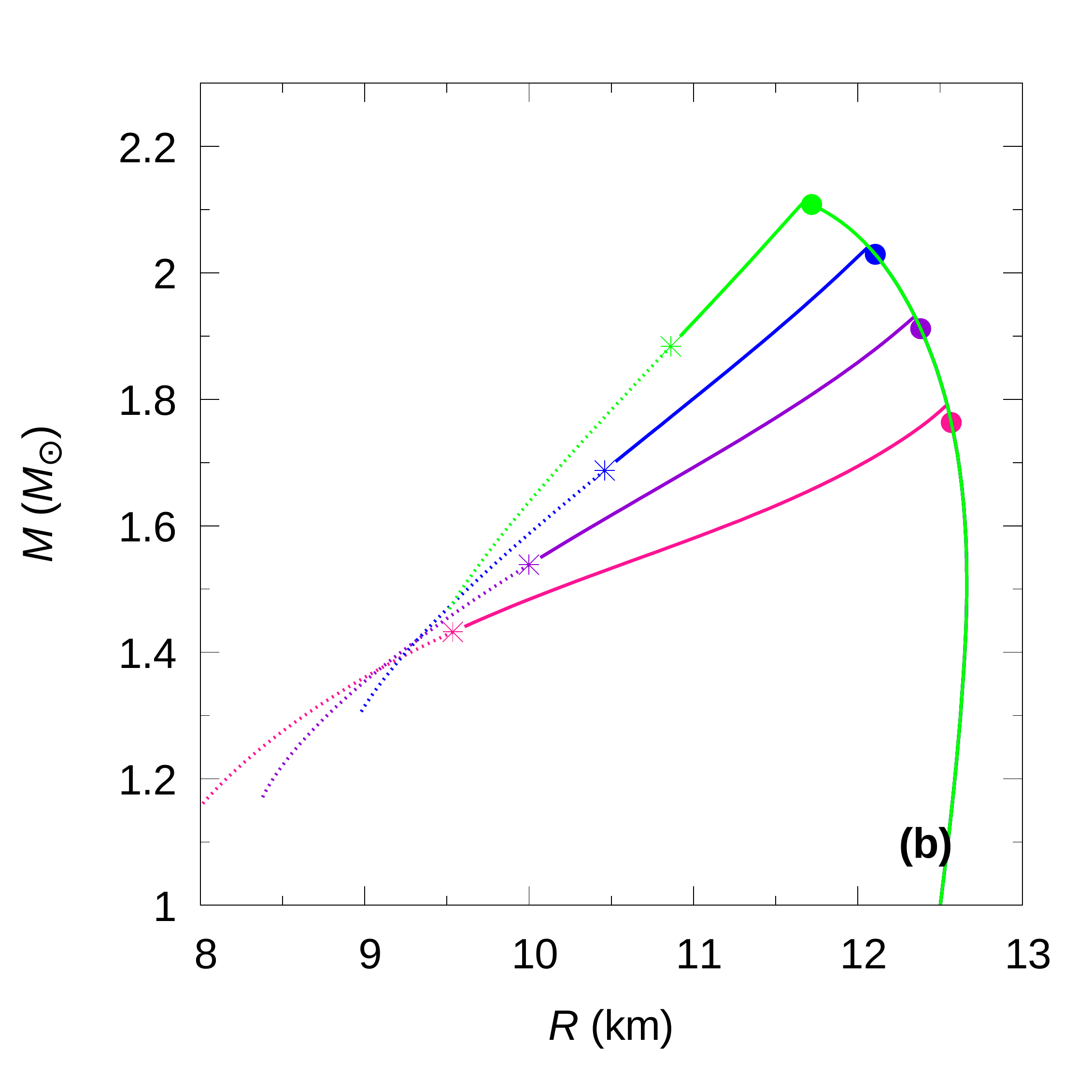}
  \caption{{ Hybrid EoS} (panel (\textbf{a})) and mass-radius
   relationship (panel (\textbf{b})) for the CFL phase at fixed gap value of
   $\Delta=35$~MeV, for~different values of the $K_{\rm v}$ parameter.
   In panel (\textbf{a}), the~grey region shows the constraints presented
     in~\cite{Annala:2019puf}.  The~solid dots in panel (\textbf{b}) indicate
   the appearance of the color superconducting phase, just before the
   maximum mass peak. For~rapid conversions, the~stellar
   configurations to the left of each maximum mass are unstable and
   the existence of HSs is only marginal. For~slow conversions, an~   extended stability branch exists. The~stable configurations are
   shown by continuous lines. The~terminal configurations are marked
   with asterisks.}
  \label{fig:cfl_delta35}
\end{figure}

\begin{figure}
  \includegraphics[width=0.9\columnwidth]{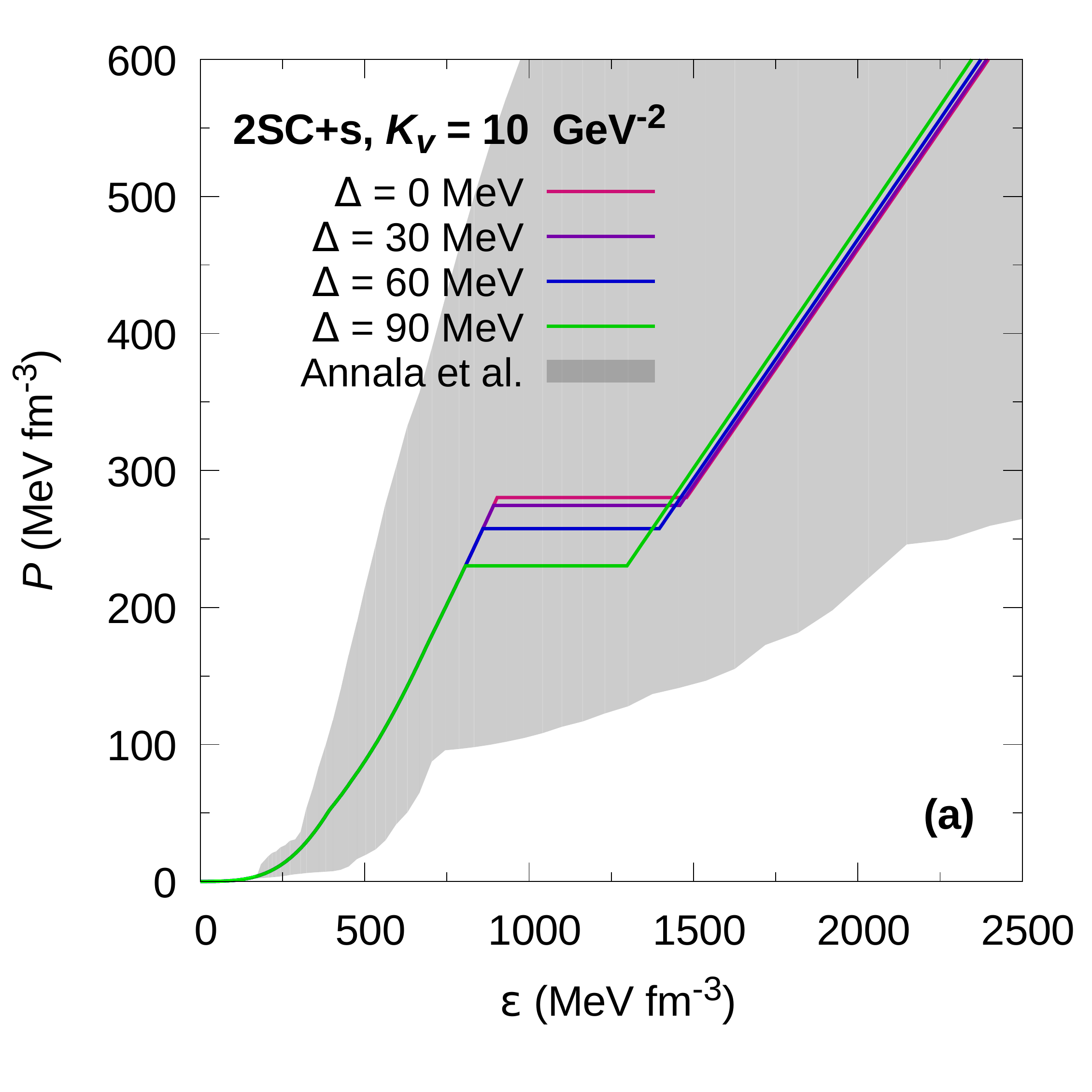}
  \includegraphics[width=0.9\columnwidth]{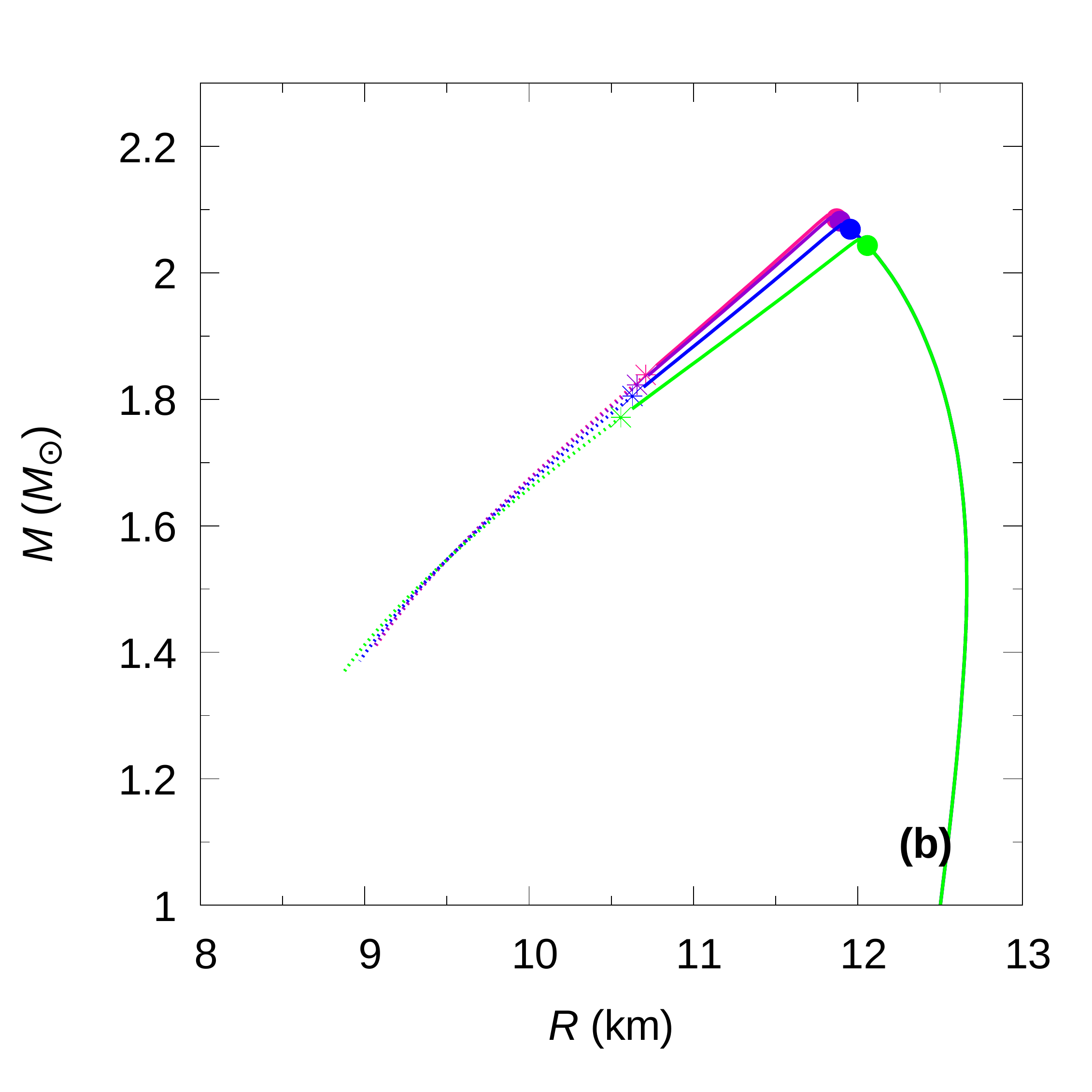}
  \caption{{Hybrid EoS} (panel (\textbf{a})) and mass-radius
   relationship (panel (\textbf{b})) for the 2SC+s phase at fixed gap value of
    $K_{\rm v} =10$~GeV$^{-2}$, for~different values of the $\Delta$ parameter.
   In panel (\textbf{a}), the~grey region shows the constraints presented
     in~\cite{Annala:2019puf}.  The~solid dots in panel (\textbf{b}) indicate
   the appearance of the color superconducting phase, just before the
   maximum mass peak. For~rapid conversions, the~stellar
   configurations to the left of each maximum mass are unstable and
   the existence of HSs is only marginal. For~slow conversions, an~   extended stability branch exists. The~stable configurations are
   shown by continuous lines. The~terminal configurations are marked
   with asterisks.}
  \label{fig:2sc_kv10}
\end{figure}

\begin{figure}
 \includegraphics[width=0.9\columnwidth]{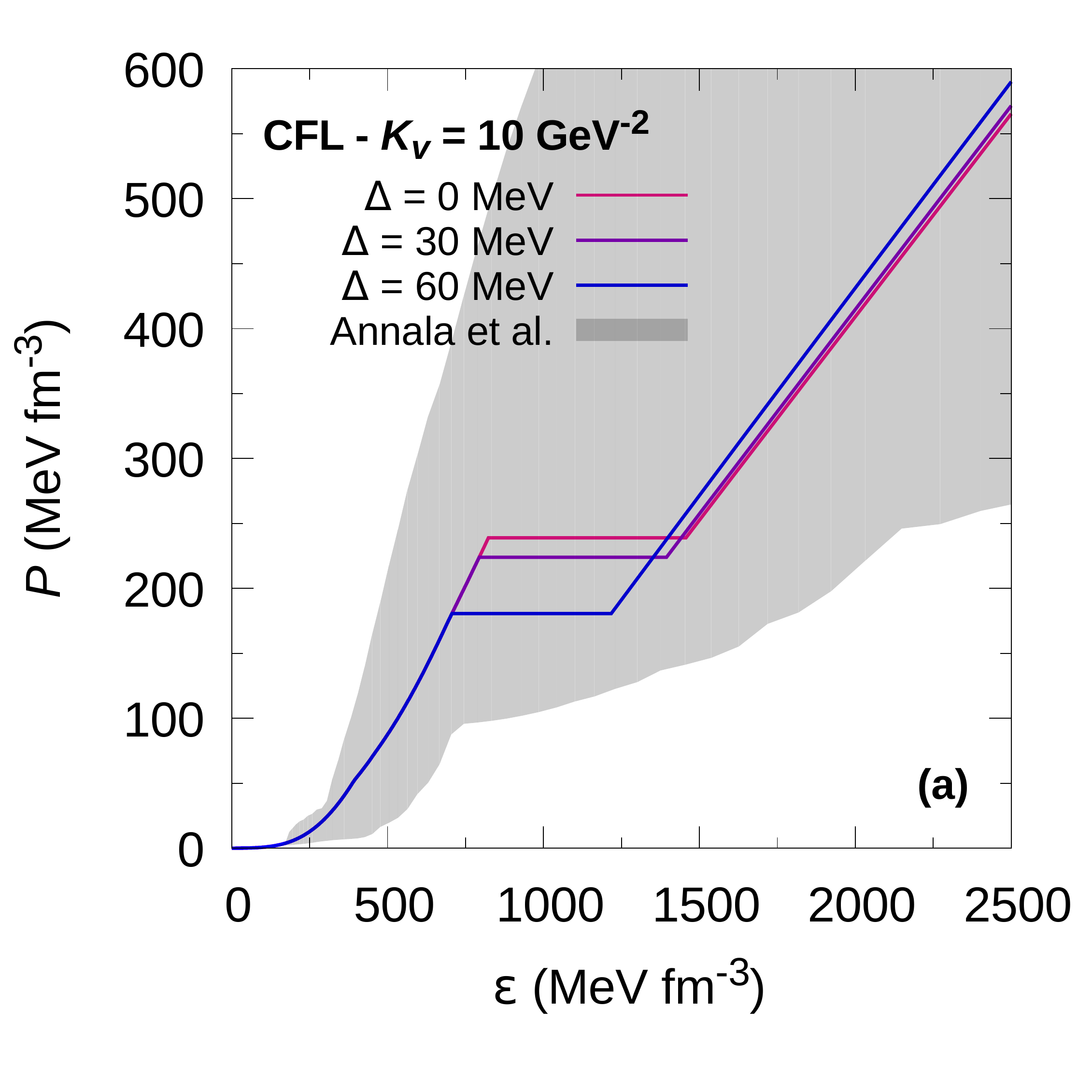}
 \includegraphics[width=0.9\columnwidth]{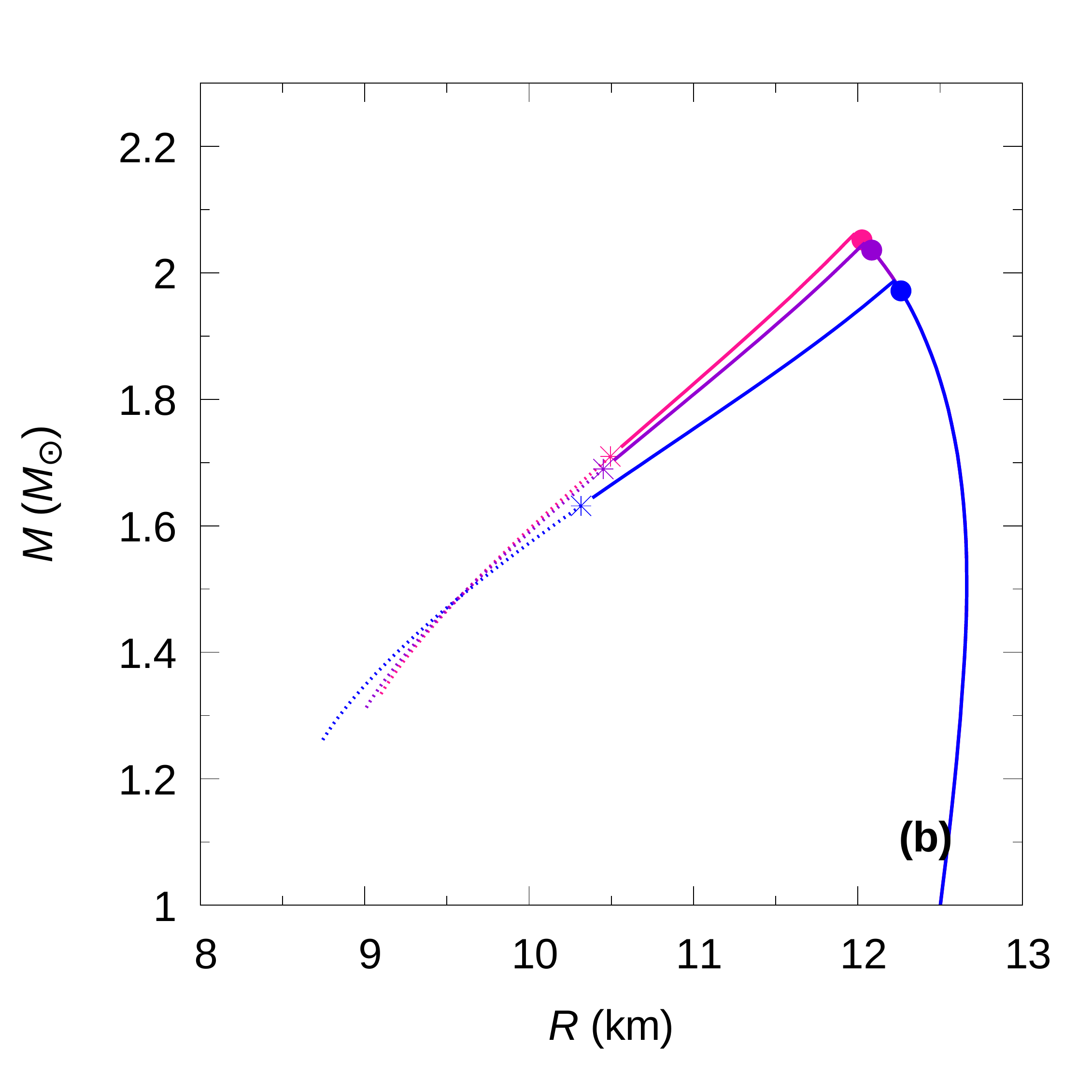}
 \caption{{(Hybrid EoS)} (panel (\textbf{a})) and mass-radius
   relationship (panel (\textbf{b})) for the CFL phase at fixed gap value of
    $K_{\rm v} =10$~GeV$^{-2}$, for~different values of the $\Delta$ parameter.
   In panel (\textbf{a}), the~grey region shows the constraints presented
     in~\cite{Annala:2019puf}.  The~solid dots in panel (\textbf{b}) indicate
   the appearance of the color superconducting phase, just before the
   maximum mass peak. For~rapid conversions, the~stellar
   configurations to the left of each maximum mass are unstable and
   the existence of HSs is only marginal. For~slow conversions, an~   extended stability branch exists. The~stable configurations are
   shown by continuous lines. The~terminal configurations are marked
   with asterisks.}
 \label{fig:cfl_kv10}
\end{figure}

We start by analyzing the effects of varying $K_{\rm v}$ and $\Delta$
values on the EoSs and the mass-radius relationship ($M$--$R$) of HSs
shown in Figures~\ref{fig:2sc_delta35}--\ref{fig:cfl_kv10}. All the hybrid EoSs shown
in these figures satisfy the constraints presented in
~\cite{Annala:2019puf}; besides, these EoSs have one common
characteristic feature, namely that the hadron-quark transition
pressure must be larger than about 200~MeV~fm$^{-3}$ so that the $2\,
M_{\odot}$-mass constraint condition can be~satisfied.

\begin{figure}
  \includegraphics[width=1.0\columnwidth]{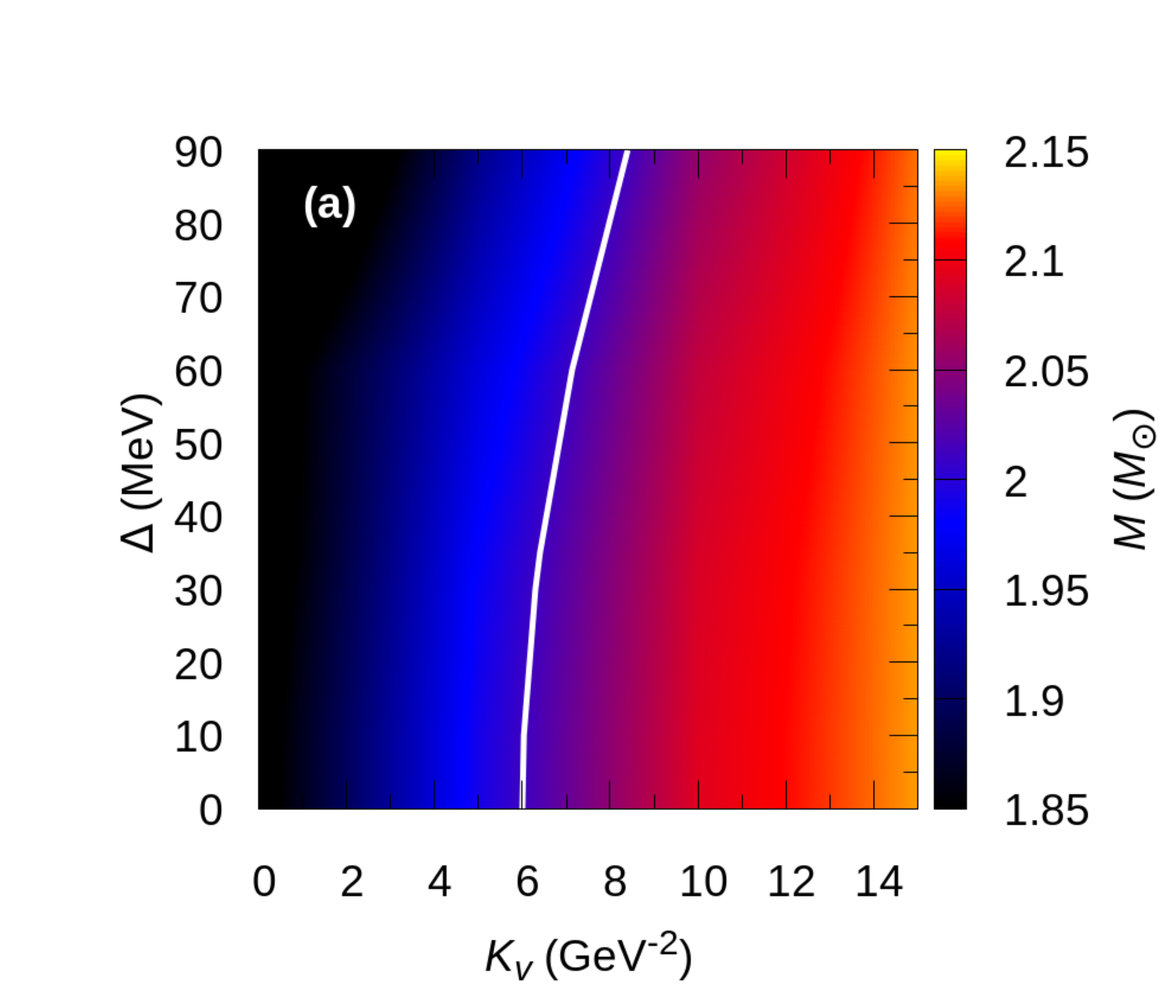}
    \includegraphics[width=1.0\columnwidth]{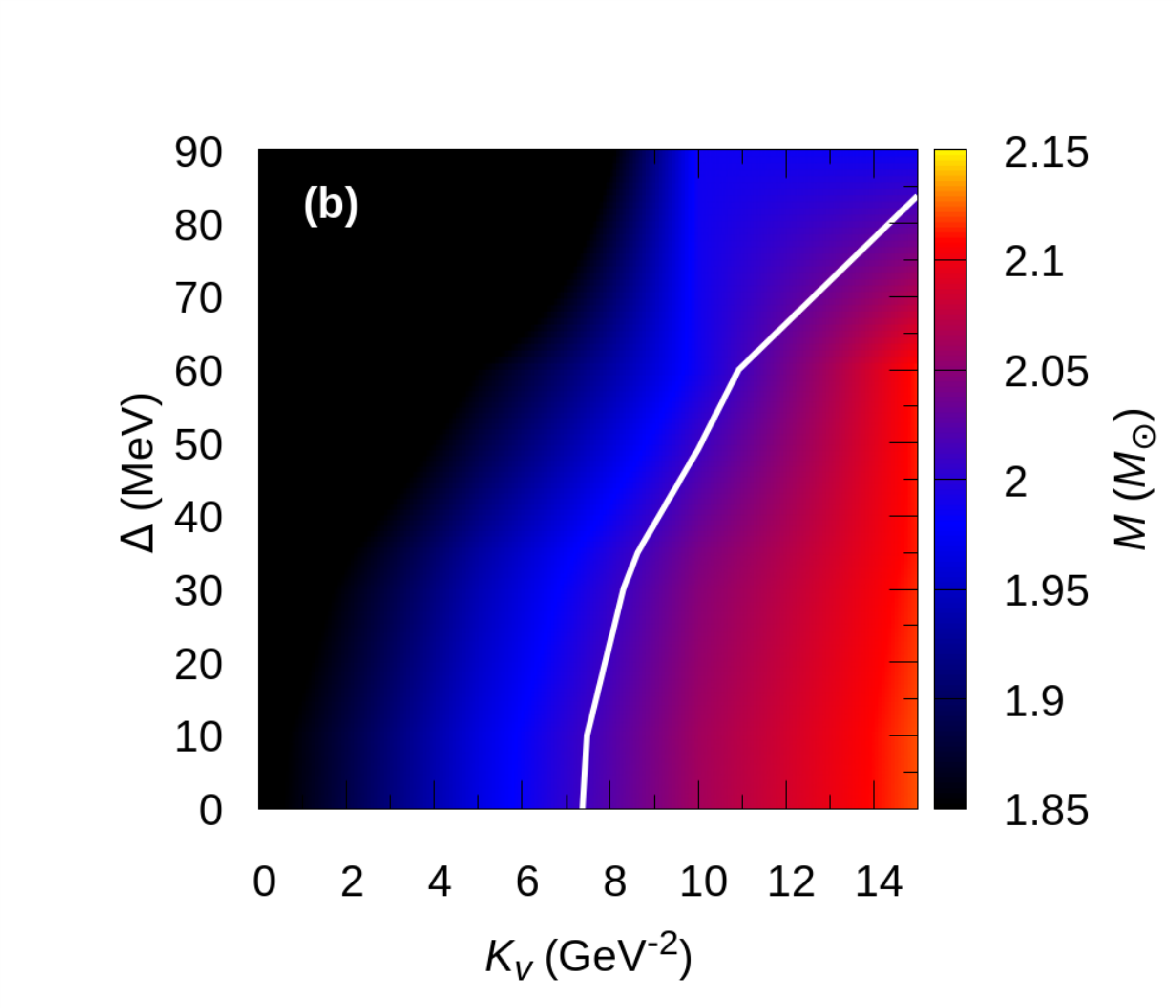}
  \caption{({Text}) Maximum mass of stars as a function of
    $\Delta$ and $K_{\rm v}$, for~the 2SC+s quark matter (panel (\textbf{a})) and
    CFL quark matter (panel (\textbf{b})). The~white curve marks the maximum
    mass constraint $M_{\rm max}=2.01M_\odot$.}
  \label{fig:mmax}
\end{figure}

\begin{figure}
  \includegraphics[width=0.9\columnwidth]{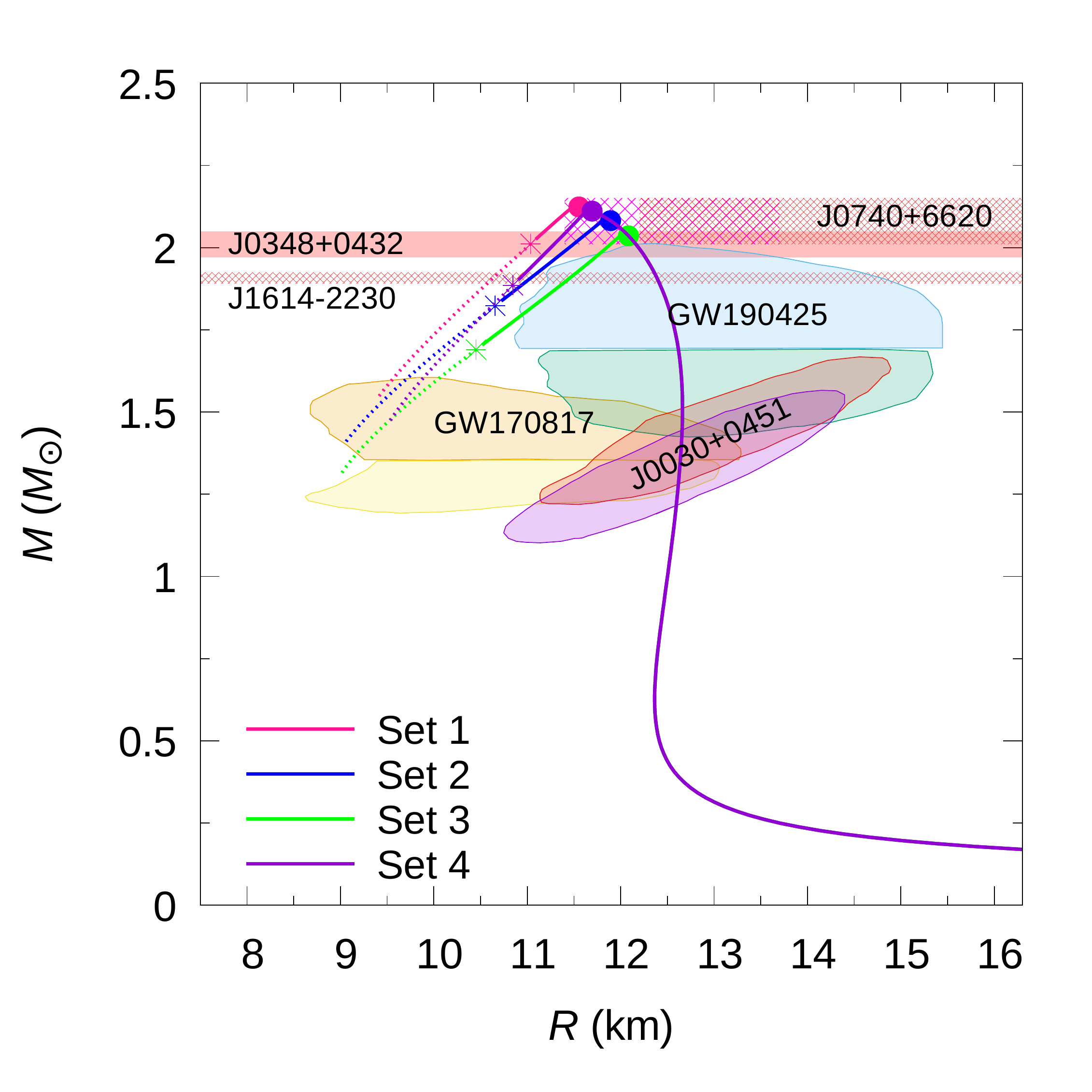}
  \caption{(Color online)  $M$--$R$ relationship of the selected EoSs
    (Table~\ref{table:set}) of this work. The~solid dots indicate the
    appearance of color superconducting quark matter in HSs, which
    happens just before the maximum-mass peaks are reached. For~a
    rapid conversion of matter, the~stellar configurations to the left
    of each maximum-mass star are unstable. For~slow conversions there
    exist extended branches of stable stars which end at the locations
    marked with asterisks. The~shaded regions (clouds) correspond to
    constraints imposed by GW170817, GW190425, and~NICER observations
    of PSR J0030+0451. The~horizontal pink stripped bands, indicate
    constraints imposed by pulsars J0740+6620, J0348+0432, and~J1614-2230.} 
  \label{fig:mraio_constraint}
\end{figure}

In Figure~\ref{fig:mraio_constraint}, we show the M-R curves that
correspond to hybrid star configurations constructed with EoSs whose
parameters are listed in Table~\ref{table:set}. These curves are
consistent with the $2\,M_\odot$ mass constraint set by massive
pulsars, NICER observations, as~well as the NS data extracted from the
gravitational-wave event GW170817 and GW190425. We see that, when
assuming slow hadron-quark conversion, each model predicts the
existence of high-mass twin stars. And~because of this possibility,
the observed $2\,M_\odot$ pulsars could be either NSs or HSs. The~radii of the latter could be up to 1.5~km smaller than those of the
NSs.  Furthermore, for~parameter set 3 of Table~\ref{table:set} we find
that the corresponding extended hybrid-star branch could even explain
the stellar high-mass component of the GW190425 binary system
{\cite{gw190425-detection}}.

We have also explored the possibility of sequential phase transitions
between the two different quark matter EoSs, i.e.,~the occurrence of a
transition of quark matter from the 2SC+s to the CFL phase. We~find that such a sequential transition is possible, but~the  $M$--$R$
relationships do not fulfill $2\, M_{\odot}$ mass constraint. The~main reasons for this is a low speed of sound of $c_s^2/c^2 \sim 0.33$ in
the extended FCM EoS and a high phase transition~pressure.

\begin{figure}
    \includegraphics[width=0.9\columnwidth]{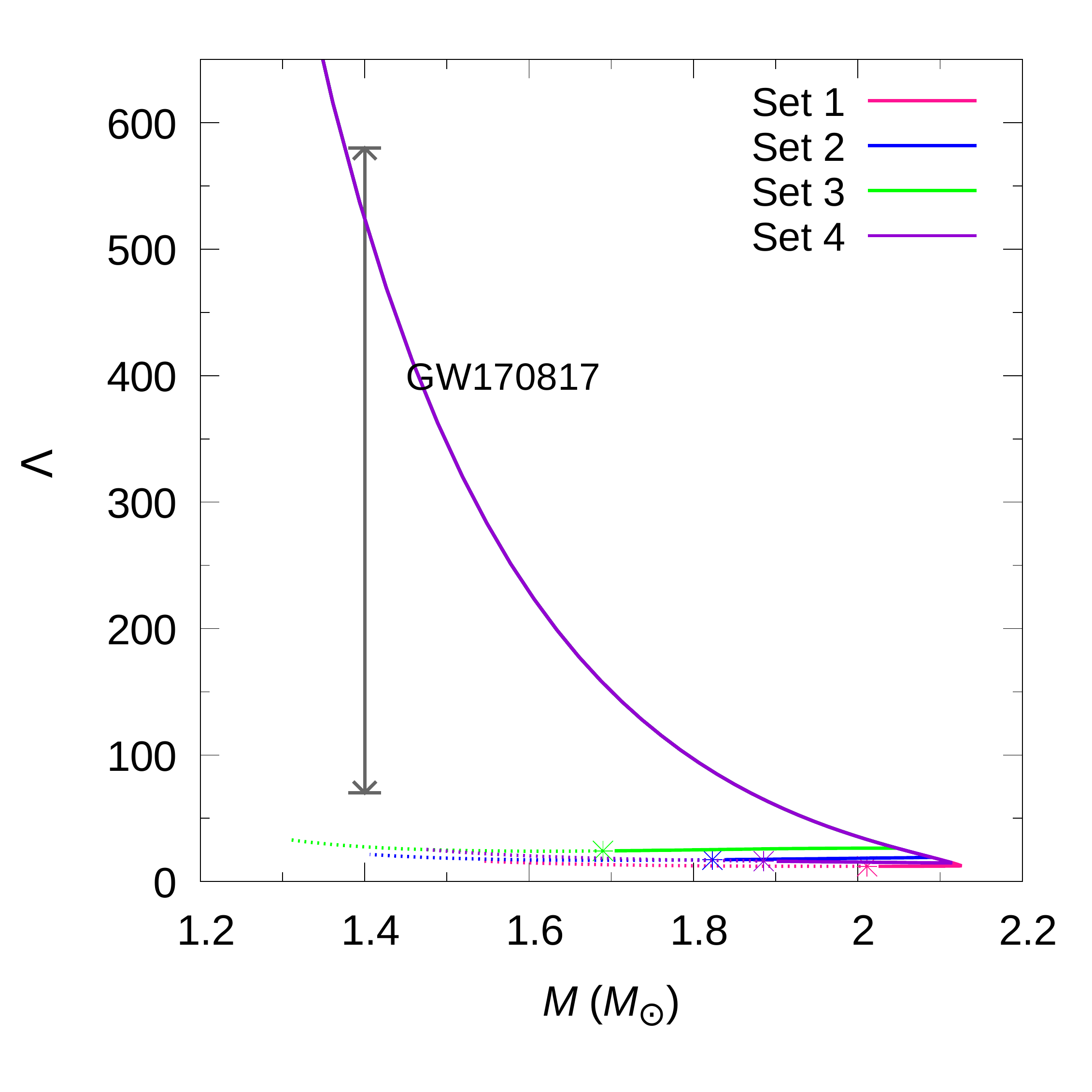}
  \caption{(Color online) Dimensionless tidal deformability as a
    function of gravitational mass, with~the constraint obtained from
    {GW170817}~\cite{Abbott:2018wiz} \textbf{me queda la duda en este tmb}. Stable stellar configurations
    beyond the maximum mass have very small values of $\Lambda$, which
    are almost independent of mass. The~positions of the terminal
    stars of the twin HSs branch (obtained for slow hadron-quark
    conversion) are marked with~asterisks.} 
  \label{fig:tidal_masa}
\end{figure}

In Figure~\ref{fig:tidal_masa}, we present the dimensionless tidal
deformability, $\Lambda$, as~a function of gravitational mass for the
stellar hybrid configurations of Figure~\ref{fig:mraio_constraint}.  {All models present} pure hadronic stars for masses $M~\leq ~1.4~M_{\odot}$, and~are consistent with the  $\Lambda _{1.4} \sim 500$ constraint deduced from GW170817. One also sees that the HSs along the twin
stellar branch have tidal deformabilities that lie on an almost
straight horizontal line.  This opens up the possibility that future
observations of NS mergers may help to shed light on the actual
existence of twin stars and hence on the behavior of matter in the
inner cores of compact~objects.  

\begin{figure}
  \includegraphics[width=0.7\columnwidth]{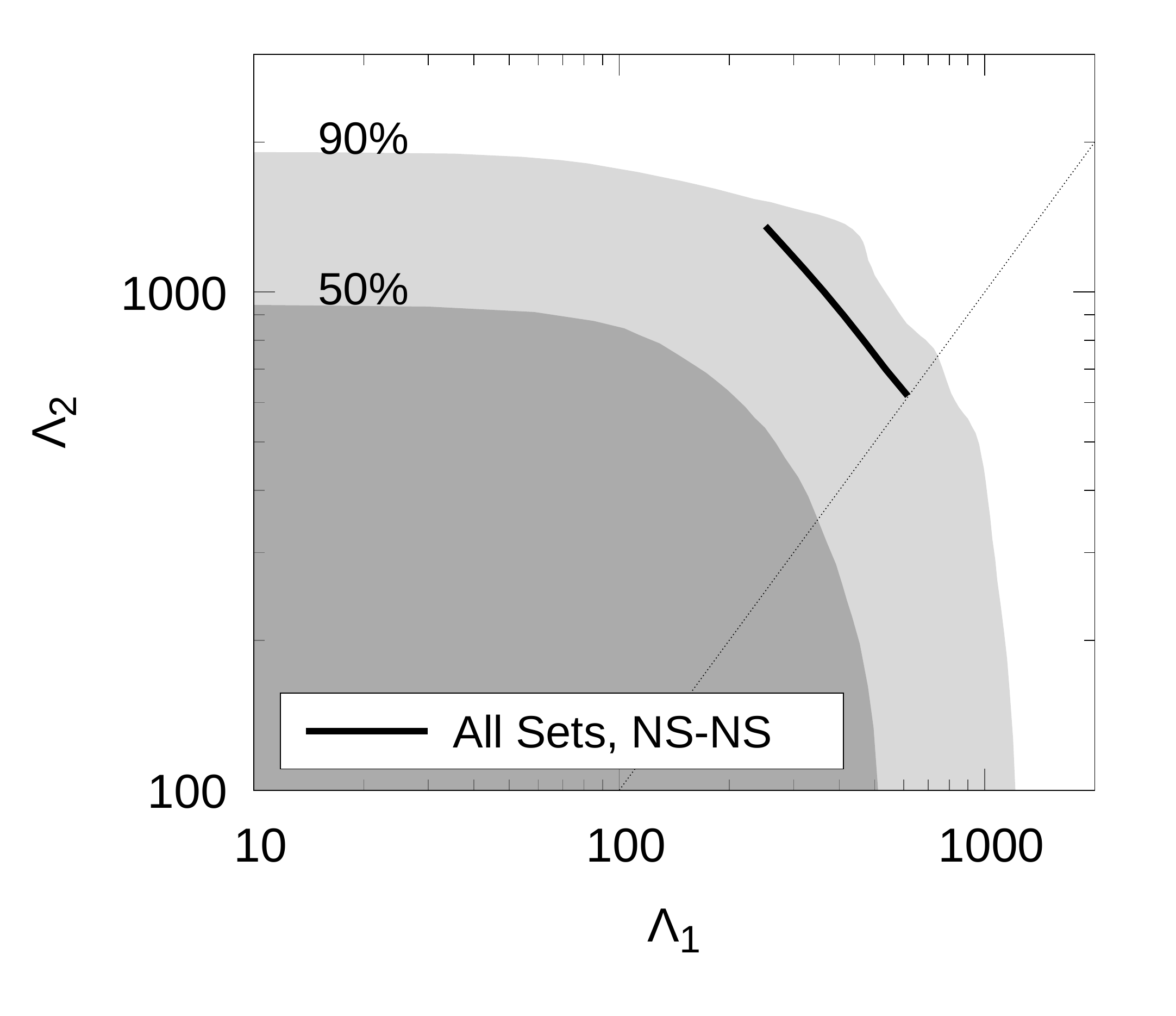}
  \caption{{Text}  Dimensionless tidal deformabilities
    $\Lambda_1$ and $\Lambda_2$ for the selected EoSs. The~solid black
    line represents the results obtained for a purely hadronic NS-NS
    merger with masses consistent with data from GW170817. The~dark
    (light) gray areas represent the 50\% (90\%) confidence limit of
    the probability contour of GW170817 and the dotted line
    corresponds to $\Lambda _1 = \Lambda _2$.}
  \label{fig:tidales}
\end{figure}

In Figure~\ref{fig:tidales}, we show the individual dimensionless tidal
deformabilities of the hybrid configurations consistent with the
observational constraints obtained after GW170817 and its
electromagnetic counterpart. The~black line represents the situation
in which the two merging objects are purely hadronic~NSs.

\section{Summary and~Conclusions}\label{sec:conclu}

In this work, we have studied hybrid EoSs and the structure of HSs
considering the effects of color superconductivity (2SC+s and CFL
phases) and vector interactions in quark matter in the framework of
the FCM model. Both color superconductivity and vector interactions
were included in the model in a phenomenological way, taking advantage
of the similarity of the FCM with the MIT bag model at the~zero temperature limit. For~the description of the hadronic phase, we used
the SW4L parametrization of the RMF model. We have assumed a sharp
hadron-quark phase transition and considered the the implications of
slow versus rapid conversions of matter at the hadron-quark
interface. This assumption dramatically modifies the traditional
picture of stability in the $M$--$R$ diagram of HSs. {For instance, if~we consider} a soft Gibbs phase transition, instead of a
  Maxwell sharp transition, the~extended stable branch does not
  exist. Since in the Gibbs case the EoS has no discontinuity or
  jump, the~stability criterion for  hybrid stars is the
  traditional one, in~which case $\partial M/\partial \epsilon _c < 0$ indicates
  unstable configurations and the maximum mass configuration is the
  last stable star in the mass-radius diagrams.  


After extending the parameter space of the FCM model, we performed a
systematic analysis of the parameters of this new space.  The~goal was
to find out whether the parameters lead to equations of state that are
consistent with present astrophysical observations.  For~this purpose
we investigated the dependencies of the EoS and the $M$--$R$
relationship of compact stars on the $K_{\rm v}$ and $\Delta$
parameters, which are related to vector interactions and color
superconductivity of the extended FCM model.  Fixed values were
assumed for the other two parameters, $V_1$ and $G_2$, of~the model.
As our investigations show, the~hybrid EoSs we determined successfully
satisfy the constraints set by \citet{Annala:2019puf} and by PSR
J1614-2230, PSR J0348+0432, PSR J0740+6620, GW170817, GW190425, and~NICER~observations.

In addition, using a specific FCM model parameter set, we have shown
that the inclusion of vector interactions and color superconductivity
plays a central role in satisfying the mass constraint set by massive
pulsars.  Specifically, we found that increasing $K_{\rm v}$ leads to a
stiffer hybrid EoS, which increases the maximum stellar mass.
However, this increase leads to shorter stability branches for the
twin stars. In~contrast, an~increase in the $\Delta$ parameter leads
to softer EoSs, both for the 2SC+s and CFL phase, which lowers the
maximum mass but leads to extended branches of stellar stability. In~general, changes in the value of $K_{\rm v}$ have a more pronounced effect
on the system properties than changes in $\Delta$. An~exception is the
CFL phase, where changes in $\Delta$ dominate the mass-radius
relationship.

We have also explored the possibility of a sequential phase transition
in HSs. We have found that although that possibility exists, the~hybrid configurations obtained from these EoSs do not satisfy the
restrictions imposed by massive pulsars. This is due to a low speed of
sound in the quark phase and a high value of the hadron-quark
transition pressure. It is worth a short discussion of this~point since the authors of Refs.~\cite{sequential, Rodriguez_2021} obtained
HSs with sequential phase transitions in the constant speed of sound
framework, which fulfill the $2\, M_{\odot}$ mass constraint. This was
possible by using a parametric EoS for the quark phase and by fixing
the hadron-quark phase transition pressure (at $p_t \sim 100$~MeV
fm$^{-3}$) as well as the quark-quark phase transition pressure (at
$p_t \sim 250$~MeV~fm$^{-3}$). Furthermore, a~high value of the speed
of sound in quark matter phases ($c_s^2/c^2 \gtrsim 0.7$) was assumed
in that~paper.

Massive HSs with sequential phase transitions were also obtained with
Nambu-Jona-Lasinio type models of quark matter~\cite{Pagliara_2008,Bonanno_2012}. However, extra ingredients are needed in these models to achieve a hadron-quark followed by a
quark-quark phase transition, because~in these models $c_s^2/c^2 \sim
0.33$.  Besides~that, an~effective bag can be added to the model to
lower the transition pressure, considering a large diquark coupling
~\cite{Pagliara_2008}. In~these works, the~transition pressures (i.e.,
$p_t \sim 40$ to 60~MeV~fm$^{-3}$ for the hadron-quark transition and
$p_t \sim100$ to 130~MeV~fm$^{-3}$ for the quark-quark transition) are
lower than in Ref.~\cite{sequential}.  In~a recent study~\cite{ferreira2020quark} it was shown that in NJL models a higher speed of sound can be achieved through the incorporation of
higher-order repulsive interactions. This affects the size of the
quark core in HSs, leading to massive hybrid configurations with
extended cores of quark matter in the rapid conversion~scenario.

Regarding the tidal deformability results, our models satisfy the
GW170817 constraint of a $1.4\, M_\odot$ star. Also the restrictions
coming from the constraints in the $\Lambda_1$--$\Lambda_2$ plane are
fulfilled. In~this case, the~purely hadronic branch already lies, for~all four parameters sets of Table~\ref{table:set}, inside the
confidence region. Remarkably, the~slow hadron-quark conversion
scenario, which leads to new stable hybrid-star branches, helps to
satisfy astrophysical constraints (similar conclusions have already
been presented in Refs.~\cite{Mariani:2019, Malfatti2020PRD,
  Rodriguez_2021}).

  
\vspace{30pt}

  \textbf{Author contributions:}
  The authors contributed equally to the  theoretical and numerical aspects of the work presented in this paper. All authors have read and agreed to the published version of the~manuscript.
  
  \vspace{6pt}

\textbf{Funding:}
D.C. is a fellow of UNLP. D.C., I.F.R.-S., M.M. and
  M.G.O. thank CONICET and UNLP (Argentina) for financial support under
  grants PIP-0714 and G157, G007. I.F.R.-S. is also partially
  supported by PICT 2019-0366 from ANPCyT (Argentina). I.F.R.-S.,
  M.G.O. and F.W. are supported by the National Science Foundation
  (USA) under Grants PHY-2012152.

    \vspace{6pt}

\textbf{Institutional review:}
Not applicable.

  \vspace{6pt}

\textbf{Informed consent:}
Not applicable.

  \vspace{6pt}

\textbf{Acknowledgments:}
The authors would like to thank V. Dexheimer and
  R. Negreiros for inviting us to submit a contribution to the Special
  Issue of Universe entitled ``Properties and Dynamics of Neutron Stars
  and Proto-Neutron~Stars''.
  
    \vspace{6pt}

  \textbf{Conflict of interest:}
The authors declare no conflict of~interest.
    \vspace{6pt}

\end{document}